\documentclass[preprint,nofootinbib,aps,superscriptaddress,tightenlines,eqsecnum]{revtex4-1}
\pdfoutput=1
\usepackage[colorlinks=true,citecolor=blue,linkcolor=purple,breaklinks=true]{hyperref}
\usepackage{graphicx}
\usepackage{mathrsfs}
\usepackage{booktabs}
\usepackage{enumerate}
\usepackage{slashed}
\usepackage{latexsym, amssymb} 
\usepackage{amsmath,subfigure,xcolor}
\usepackage{multirow}

\newcommand{\vev}{{\textit{vev}} }
\newcommand{\vevs}{{\textit{vev}}s }
\newcommand{\tr}{{\rm{Tr}}}

\newcommand{\beq}{\begin{equation}}
\newcommand{\eeq}{\end{equation}}
\newcommand{\bea}{\begin{eqnarray}}
\newcommand{\eea}{\end{eqnarray}}

\newcommand{\nn}{\nonumber}

\def\a{\alpha}

\def\b{\beta}
\def\k{\kappa}

\def\l{\lambda}
\def\m{\mu}

\def\s{\sigma}

\begin{document}
\title{Domain wall constraints on the doublet left-right symmetric model from pulsar timing array data}
\author{Dhruv Ringe\footnote{phd1901151004@iiti.ac.in}}
\affiliation{{\small Department of Physics, Indian Institute of Technology Indore, \\Khandwa Road, Simrol, Indore, Madhya Pradesh-453552, India}}

\begin{abstract}
{\noindent Recent evidence of a stochastic gravitational wave (GW) background found by NANOGrav and other pulsar timing array (PTA) collaborations has inspired many studies looking for possible sources. We consider the hypothesis that the GW signature is produced by domain walls  (DWs) arising in the doublet left-right symmetric model (DLRSM) due to the spontaneous breaking of the discrete parity symmetry. The DW network consists of two types of DWs, namely $Z_2$ and $LR$ DWs, which have different surface tensions. We find kink solutions for both types of DWs and obtain the parametric dependence of the surface tension. Considering the GW signal from the DLRSM DW model with and without the contribution from supermassive black hole binaries, we perform a Bayesian analysis using the PTA data to estimate the posterior distribution and identify best-fit parameter ranges. The PTA data favors a parity-breaking scale of $\mathcal{O}(10^5)$\,GeV, and a biased potential $V_{\rm{bias}}\sim (\mathcal{O}(100)\,\rm{MeV})^4$. The model with only DLRSM DWs is slightly favored over the model where additional SMBHB contribution is considered. }
\end{abstract}

\maketitle
%
\section{Introduction}
The 15-year dataset (NG15) of the NANOGrav collaboration\,\cite{NANOGrav:2023gor} shows compelling evidence of a stochastic gravitational wave background (SGWB) at nanoHertz frequencies. This evidence has been corroborated at varying significance levels by other pulsar timing arrays (PTAs) such as the European Pulsar Timing Array (EPTA) in collaboration with the Indian Pulsar Timing Array (InPTA)\,\cite{EPTA:2023fyk}, Parkes Pulsar Timing Array (PPTA)\,\cite{Reardon:2023gzh}, which are all part of the International Pulsar Timing Array (IPTA) consortium. While more data is needed to claim a discovery, discussing the possible sources of such a SGWB is interesting. The standard astrophysical interpretation, of the SGWB produced by in-spiralling supermassive black hole binaries (SMBHBs) scattered across the universe, is in slight tension with the data\,\cite{NANOGrav:2023gor,Ellis:2023dgf,Ghoshal:2023fhh,Shen:2023pan,Broadhurst:2023tus,Bi:2023tib,Zhang:2023lzt}. Other explanations of cosmological origin have been discussed in the literature, including the gravitational waves (GWs) from the density perturbations after inflation\,\cite{Franciolini:2023pbf,Vagnozzi:2023lwo,Inomata:2023zup,Ebadi:2023xhq,Liu:2023ymk,Abe:2023yrw,Unal:2023srk,Firouzjahi:2023lzg,Bari:2023rcw,Cheung:2023ihl,Bhaumik:2023wmw,Gorji:2023sil}, first-order phase transitions\,\cite{Fujikura:2023lkn,Addazi:2023jvg,Bai:2023cqj,Megias:2023kiy,Han:2023olf,Zu:2023olm,Ghosh:2023aum,DiBari:2023upq,Cruz:2023lnq,Gouttenoire:2023bqy,Ahmadvand:2023lpp,An:2023jxf,Wang:2023bbc}, and topological defects such as domain walls (DWs)\,\cite{Ferreira:2022zzo,Kitajima:2023cek,Guo:2023hyp,Blasi:2023sej,Gouttenoire:2023ftk,Barman:2023fad,Lu:2023mcz,Li:2023tdx,Du:2023qvj,Gelmini:2023kvo,Zhang:2023nrs} and cosmic strings\,\cite{Ellis:2023tsl,Kitajima:2023vre,Wang:2023len,Lazarides:2023ksx,Eichhorn:2023gat,Servant:2023mwt,Antusch:2023zjk,Fu:2024rsm,Yamada:2023thl,Ge:2023rce}. 

Comparative analyses of the possible SGWB sources reveal that many of these models provide a better fit compared to the standard SMBHB interpretation\,\cite{NANOGrav:2023hvm,Ellis:2023oxs,Wu:2023hsa}. In this paper, we consider DWs as the possible source of the signal, for which the Bayes factor is $\mathcal{O}(10)$ when compared to the fiducial SMBHB model\,\cite{NANOGrav:2023hvm}. From the particle physics perspective, DWs are formed when a discrete symmetry is spontaneously broken. The GW spectrum from DWs depends on the surface tension, $\sigma$, of the walls, and the bias potential, $V_{\rm{bias}}$. The PTA data is compatible with DWs for values roughly, $\sigma\sim (100\,\rm{TeV})^3$, and the bias, $V_{\rm{bias}}\sim (100\,\rm{MeV})^4$\,\cite{Ferreira:2022zzo}. A microscopic model for DWs at the electroweak (EW) scale, $v_{\rm{EW}}\sim 246.02$\,GeV, cannot yield DWs with such a high surface tension, and thus a viable microscopic DW model must incorporate high-scale physics. Given that the EW interactions in the standard model (SM) maximally violate parity, high-scale extensions of SM that respect the parity symmetry, $\mathcal{P}$, provide an interesting way to generate DWs with sufficiently large surface tension.

Left-right symmetric models\,(LRSMs)\,\cite{Pati:1974yy,Mohapatra:1974gc,PhysRevD.11.566,Senjanovic:1975rk,Senjanovic:1978ev} are well-motivated extensions of the standard model (SM) where the gauge group is extended from $\mathcal{G}_{\rm{SM}} = SU(3)_c\times SU(2)_L\times U(1)_{Y}$ to $\mathcal{G}_{\rm{LRSM}} = SU(3)_c\times SU(2)_L\times SU(2)_R\times U(1)_{B-L}$. An additional discrete $\mathcal{P}$ symmetry can be easily incorporated into $\mathcal{G}_{\rm{LRSM}}$, allowing for the possibility of DW formation. The various realizations of LRSM differ from each other, depending on the scalars involved in the spontaneous breaking of $\mathcal{G}_{\textrm{LRSM}}$ to $\mathcal{G}_{\rm{SM}}$. They also differ in the mechanism of fermion mass generation. A widely studied realization is the triplet LRSM (TLRSM), where the scalar sector involves two triplets and a bidoublet\,\cite{Maiezza:2016ybz,PhysRevD.44.837,Senjanovic:2016bya}. When the scalar sector contains two doublets and a bidoublet, it is called the doublet LRSM (DLRSM)\,\cite{Senjanovic:1978ev,Mohapatra:1977be}. Other versions of LRSM are also studied in the literature \,\cite{Ma:1986we,Babu:1987kp,Ma:2010us,Frank:2019nid,Frank:2020odd,Graf:2021xku}. It was recently shown that the pattern of electroweak symmetry breaking (EWSB) in DLRSM can be quite different from the other versions of LRSM, with interesting consequences from precision observables\,\cite{Bernard:2020cyi} and Higgs data\,\cite{Karmakar:2022iip}. 

In this paper, we study the DWs arising in parity-symmetric DLRSM. Previous discussions on DWs in LRSM\,\cite{Yajnik:1998sw,Borah:2022wdy,Chakrabortty:2019fov,Mishra:2009mk,Borah:2011qq,Banerjee:2020zxw,Borboruah:2022eex,Barman:2023fad,Banerjee:2023hcx}  mainly centered around TLRSM. In refs.\,\cite{Yajnik:1998sw,Borboruah:2022eex}, the kink solutions for `left-right' ($LR$) DWs were presented for a few benchmark points. The GW signature was studied in refs.\,\cite{Borah:2022wdy,Barman:2023fad}, where the benchmarks were chosen based on an approximate dependence of the DW surface tension without explicitly solving the kink equations. In this paper, we show that two types of DWs are formed in DLRSM, namely, $Z_2$ and $LR$ DWs, with different surface tensions. For both types, we solve the kink equations to obtain the parametric dependence of the DW surface tension and show that the $Z_2$ DWs are unstable. We then obtain the GW signature in terms of the DLRSM parameters. While qualitative similarity is expected in the GW signature of DWs
from DLRSM and TLRSM, we perform a Bayesian analysis on the PTA data to constrain the model parameters. We consider the GW spectrum from DLRSM DWs with and without the contribution from SMBHBs. The discussion of this paper can be easily carried over to TLRSM.   
  
In Sec.\,\ref{Sec:Model}, we briefly discuss the scalar potential of the $\mathcal{P}$-symmetric DLRSM. In Sec.\,\ref{sec: DW}, we discuss the vacuum structure of the DLRSM effective potential and study the DW solutions. In Sec.\,\ref{sec: GW}, we discuss the GW spectrum resulting from DLRSM DWs. We present our results in Sec.\,\ref{sec: results}, where we perform the Markov chain Monte Carlo (MCMC) analysis, and discuss the detection prospects at upcoming GW observatories. Finally, we summarize our findings and make concluding remarks in Sec.\,\ref{sec: discussion}.

\section{The model}
\label{Sec:Model}
The gauge group of parity-symmetric DLRSM is,
$$\mathcal{G}_\text{LRSM} = \mathcal{P}\times SU(3)_c\times SU(2)_L\times SU(2)_R\times U(1)_{B-L}.$$ 
For an overview of DLRSM, please refer to\,\cite{Senjanovic:1978ev,Mohapatra:1977be,Bernard:2020cyi}. The scalar sector has a complex bi-doublet $\Phi$, and two doublets $\chi_L$ and $\chi_R$. The scalar multiplets are,
\bea
\Phi = \begin{pmatrix}
    \phi_1^0 & \phi_2^+ \\
    \phi_1^- & \phi_2^0
\end{pmatrix} \sim (1,2,2,0), ~ \chi_L = \begin{pmatrix}
    \chi_L^+ \\
    \chi_L^0
\end{pmatrix} \sim (1,2,1,1), ~ \text{and}~\chi_R = \begin{pmatrix}
    \chi_R^+ \\
    \chi_R^0
\end{pmatrix} \sim (1,1,2,1),\nn\\
\eea
where the parentheses indicate the representation of the multiplets under $SU(3)_c$, $SU(3)_L$, $SU(3)_R$, and $U(1)_{B-L}$ respectively. The group $\mathcal{P}$ denotes the discrete parity symmetry under the exchange $L\leftrightarrow R$, with the action given by,
\beq
\mathcal{P}: Q_L\leftrightarrow Q_R,~l_L\leftrightarrow l_R,~\chi_L\leftrightarrow\chi_R,~\Phi\leftrightarrow\Phi^{\dagger},
\eeq
where $Q_L,~Q_R,~l_L,~l_R$ are the left and right-handed quark and lepton doublets respectively\,\cite{Bernard:2020cyi}. The $\mathcal{P}$ symmetry imposes the condition $g_L=g_R$ on the $SU(2)_L$ and $SU(2)_R$ gauge couplings. The scalar potential is given by\,\cite{Bernard:2020cyi},
\bea\label{eq: potential}
V  &=& V_{\chi} + V_{\chi \Phi} + V_{\Phi},\nn\\
V_{\chi} &=& - \m_3^2\ [\chi_L^{\dagger} \chi_L + \chi_R^{\dagger} \chi_R] + \rho_1\  [(\chi_L^{\dagger} \chi_L )^2 + (\chi_R^{\dagger} \chi_R )^2]
 + \rho_2\  \chi_L^{\dagger} \chi_L \chi_R^{\dagger}\chi_R, \nn\\ 
V_{\chi \Phi} &=& \m_4\  [\chi_L^{\dagger} \Phi \chi_R + \chi_R^{\dagger} \Phi^{\dagger} \chi_L] + \m_5\  [\chi_L^{\dagger} \tilde{\Phi} \chi_R + \chi_R^{\dagger}\tilde{\Phi}^{\dagger}\chi_L ]\nn\\
 &&+ \alpha_1\tr(\Phi^{\dagger} \Phi ) [\chi_L^{\dagger}\chi_L + \chi_R^{\dagger}\chi_R ]
 + \Big\{ \frac{\alpha_2}{2} \ [\chi_L^{\dagger} \chi_L  \tr(\tilde{\Phi} \Phi^{\dagger} ) + \chi_R^{\dagger} \chi_R  \tr(\tilde{\Phi}^{\dagger} \Phi )] + {\rm h.c.} \Big\} \nn\\
 &&+ \alpha_3\ [\chi_L^{\dagger}\
 \Phi \Phi^{\dagger}\chi_L + \chi_R^{\dagger} \Phi^{\dagger} \Phi  \chi_R  ] 
 + \alpha_4\ [\chi_L^{\dagger}\
 \tilde{\Phi} \tilde{\Phi}^{\dagger}\chi_L + \chi_R^{\dagger} \tilde{\Phi}^{\dagger} \tilde{\Phi}  \chi_R  ],\nn\\
V_{\Phi}  &=& -\m_1^2\tr(\Phi^{\dagger}\Phi) - \m_2^2\ [\tr(\tilde{\Phi}\Phi^{\dagger}) + \tr(\tilde{\Phi}^{\dagger} \Phi)] + \l_1[\tr(\Phi^{\dagger}\Phi)]^2\nn\\
&&  + \l_2\ [ [\tr(\tilde{\Phi} \Phi^{\dagger})]^2
 + [\tr(\tilde{\Phi}^{\dagger} \Phi)]^2 ] + \l_3\text{Tr}(\tilde{\Phi} \Phi^{\dagger}) \, \tr(\tilde{\Phi}^{\dagger} \Phi)\nn\\
 &&  + \l_4\tr(\Phi^{\dagger}\Phi) \, [\tr(\tilde{\Phi}\Phi^{\dagger})+ \tr(\tilde{\Phi}^{\dagger}\Phi)],
 \label{eq:scalarpotential}
\eea
where $\tilde{\Phi}\equiv \sigma_2\Phi^*\sigma_2$, and we take all parameters to be real. The pattern of symmetry breaking is:
$$\mathcal{P}\times SU(2)_L\times SU(2)_R \times U(1)_{B-L}\xrightarrow{~~\langle\chi_R\rangle~~} SU(2)_L\times U(1)_{Y} \xrightarrow{\langle\Phi\rangle,\langle\chi_L\rangle} U(1)_{\rm{em}} . $$The following charge-preserving and $CP$-preserving \vev structure achieves the desired symmetry-breaking pattern:
\beq
     \langle\Phi \rangle = \frac{1}{\sqrt{2}}\begin{pmatrix}
     \k_1 & 0\\
     0 & \k_2
 \end{pmatrix}, ~ \langle\chi_L\rangle = \frac{1}{\sqrt{2}} \begin{pmatrix}
     0\\
     v_L
 \end{pmatrix}, ~ \langle\chi_R \rangle = \frac{1}{\sqrt{2}}\begin{pmatrix}
     0\\
     v_R
 \end{pmatrix}\,.
 \eeq
The \vevs $\k_1,\k_2$ and $v_L$ follow the relation, $\k_1^2+\k_2^2+v_L^2 = v_{\rm{EW}}^2$, where $v_{\rm{EW}}= 246.02$\,GeV. The absence of a right-handed gauge boson in collider searches\,\cite{Solera:2023kwt} dictates the hierarchy of scales in DLRSM: $v_R\gg \k_1,\k_2,v_L$. Since the potential is parity-symmetric, an alternative hierarchy where $v_L\gg \k_1,\k_2,v_R$, is also possible but is not realized in nature. We denote the $\mathcal{P}$-symmetry breaking scale by $v_0$, such that $v_0\gg v_{\rm{EW}}$. In the next section, we discuss how spontaneous breaking of the discrete $\mathcal{P}$-symmetry gives rise to regions of disconnected vacuua separated by DWs. 

\section{Domain walls in DLRSM}\label{sec: DW}
The existence of DWs can be inferred from the minima of the effective potential in the $v_R-v_L$ plane. The tree-level effective potential of DLRSM is, 
\beq
V_{0}\equiv V_{\chi}(\langle\chi_L\rangle,\langle\chi_R\rangle) + V_{\chi \Phi}(\langle\chi_L\rangle,\langle\chi_R\rangle,\langle\Phi\rangle) + V_{\Phi}(\langle\Phi\rangle).
\eeq

The contributions to the one-loop finite temperature effective potential were discussed in ref.\,\cite{Karmakar:2023ixo}. A thorough analysis would require numerically calculating the full one-loop effective potential. Here we make some approximations to simplify the discussion. Symbolically,
\beq
V_{\rm{eff}} = V_0 + V_1 + V_{1T},
\eeq
where $V_1$ and $V_{1T}$ are the one-loop zero-temperature and finite temperature corrections respectively. The effective potential obeys the symmetry of the tree-level potential. While we can always fix the zero-temperature minima and masses at their tree-level values by adding a finite counter-term to $V_1$, the role of $V_{1T}$ should be analyzed. $V_{1T}$ has the form,
\beq
V_{1T}(\phi,T) = \sum_{i\in \rm{heavy}} n_i\frac{T^4}{2\pi^2} J_{b/f}\bigg(\frac{m^2_i(\phi)}{T^2}\bigg) + \sum_{i\in \rm{light}} n_i\frac{T^4}{2\pi^2} J_{b/f}\bigg(\frac{m^2_i(\phi)}{T^2}\bigg),
\eeq
where the sum runs over all fields, generically labeled by $\phi$. The term `heavy' denotes $v_0$-scale fields $\chi_L,\chi_R$, and the gauge bosons,\footnote{$W_L^{\pm}$ and $Z_L$ are taken as heavy fields since $v_L$ can take large values in some domains.} $W^{\pm}_{L,R},Z_{L,R}$. Similarly, `light' denotes the EW-scale fields including $\Phi$, the fermions, the photon, and gluons. $m_i$ are the field-dependent masses, and $n_i$ is the number of degrees of freedom for species $i$. The function $J_b$ ($J_f$) is defined for bosons (fermions) and has well-known high-$T$ and low-$T$ expansions\,\cite{Cline_1997}.  Write $x^2 \equiv \frac{m^2_i}{T^2}$, then for $x^2\ll 1$, 
\begin{eqnarray}
J_f(x^2) \approx &-&\frac{7\pi^4}{360} + \frac{\pi^2}{24} x^2 \label{eq: highTf}\\
J_b(x^2) \approx &-&\frac{\pi^4}{45} + \frac{\pi^2}{12} x^2 - \frac{\pi}{6}\big(x^2\big)^{3/2}.\label{eq: highTb}
\end{eqnarray}
The $x^2$-dependent term in $J_{b/f}$ leads to symmetry-restoration at high-$T$. For $x^2\gg 1$, both fermions and bosons have the same expansion with an exponential suppression due to the Boltzmann factor\,\cite{Cline_1997},\beq
J_{b/f}(x^2) \approx -\exp\bigg(-(x^2)^{1/2}\bigg)\bigg(\frac{\pi}{2}(x^2)^{3/2}\bigg)^{1/2}\label{eq: lowT}.
\eeq

DWs are formed at temperatures below the parity-breaking scale $v_0$, i.e. $T\lesssim v_0$. The network exists for a temperature range $T_{\rm{ann}}\lesssim T\lesssim v_0$, where $T_{\rm{ann}}$ is the DW annihilation temperature (Section\,\ref{sec: GW}). The DWs of DLRSM are topologically stable even after electroweak symmetry breaking (EWSB) and can survive right till the epoch of Big Bang nucleosynthesis (BBN). The evolution of the DW network before and after EWSB should therefore be considered separately.

\begin{itemize}
\item \textbf{Before EWSB}: When $v_{\rm{SM}}\ll T \lesssim v_0$, we can set the \vevs $\k_1=\k_2=0$ as the EW symmetry is restored. On the other hand, the contribution of the `heavy' fields to the one-loop temperature corrections is Boltzmann-suppressed according to eq.\,(\ref{eq: lowT}). So the effective potential in terms of the background fields $v_L$ and $v_R$ is given by,
\bea\label{eq: Veff}
    V_{\rm{eff}}(v_L,v_R) &=& V_{\chi}(v_L,v_R)\nn\\
    &=& -\frac{\mu_3^2}{2}(v_L^2+v_R^2) + \frac{\rho_1}{4}(v_L^4+v_R^4) + \frac{\rho_2}{4}v_L^2v_R^2.
\eea
The minima and saddle points of $V_{\chi}$ are given in eq.\,(\ref{eq: min}) and eq.\,(\ref{eq: saddle}) respectively.

\begin{figure}[tbp]
    \centering
    \includegraphics[width=.7\textwidth]{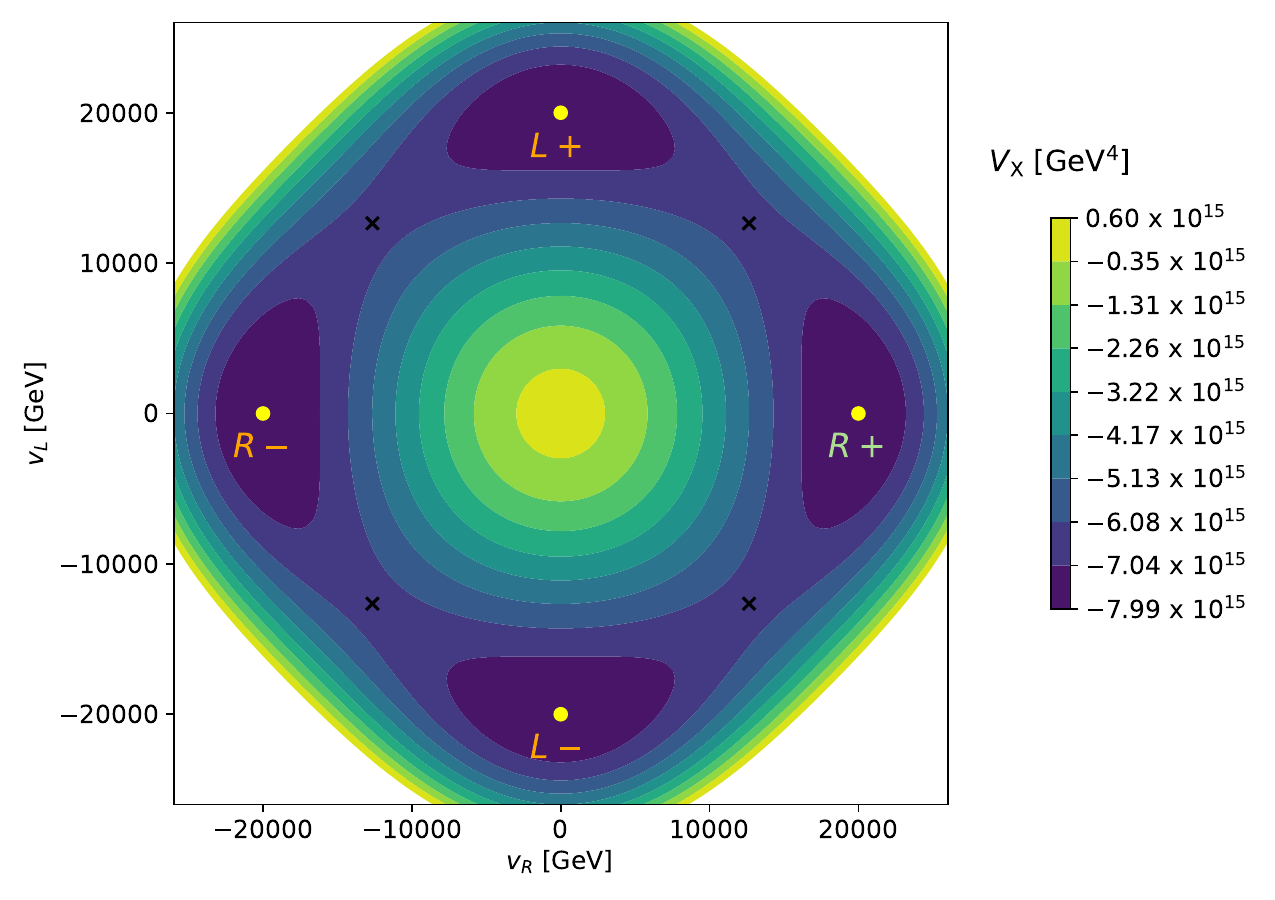}
    \caption{The potential $V_{\chi}$ for $v_0 = 20$\,TeV, $\rho_1 = 0.2$, and $\rho_2 = 0.6$. Yellow dots indicate the minima, while black crosses show the saddle points. The minimum, $R+$, shown in green, is consistent with the observed phenomenology.} 
    \label{fig: Z4}
\end{figure}

\item \textbf{After EWSB}: When $T_{\rm{ann}}\lesssim T \lesssim v_{\rm{SM}}$, the bidoublet acquires a non-zero \vev. In this regime, the thermal contributions of $v_0$-scale fields as well as EW-scale fields are Boltzmann-suppressed. The effective potential is modified as
\bea
V_{\rm{eff}}(v_L,v_R) &=& V_{\chi}(v_L,v_R) + V_{\chi\Phi}(v_L,v_R;\k_1,\k_2)\nn\\
 &=& c_1 (v_L^2+v_R^2) + c_2 (v_L^4+v_R^4) + c_3 v_L^2v_R^2 + c_4 v_Lv_R.
\eea
The $V_{\Phi}$ term has been dropped since it is independent of $v_L$ and $v_R$. The coefficients, $c_1,~c_2,~c_3,~c_4$, are given by,
\bea
&c_1 = -\frac{\mu_3^2}{2} + \frac{1}{4}[\k_1^2(\a_1+\a_4)+\k_2^2(\a_1+\a_3) + 2\k_1\k_2\a_2],~~c_2 = \frac{\rho_1}{4},&\nn \\
&c_3 = \frac{\rho_2}{4},~~c_4 = \frac{1}{\sqrt{2}}(\k_2\mu_4 + \k_1\mu_5).&\nn
\eea
Non-zero values of $\k_1$ and $\k_2$ slightly change the positions of the minima of the effective potential. We verified that the contribution of $V_{\chi\Phi}$ to the effective potential is numerically insignificant in the parameter space of interest, since $\k_1,\k_2\ll v_0$.
\end{itemize}

Hence, the DW structure is primarily governed by $V_{\chi}$, given in eq.\,(\ref{eq: Veff}). In polar coordinates, $v_R = v \cos\theta,~v_L = v \sin\theta$, eq.\,(\ref{eq: Veff}) becomes, 
\beq
V_{\chi}(v,\theta) = -\frac{\mu_3^2}{2} v^2 + \frac{v^4}{4}\Big(\rho_1 + \frac{\rho_{21}}{2}\sin^2 2\theta\Big),
\eeq
where $\rho_{21} = \rho_2/2-\rho_1$. For $\mu_3^2>0,~\rho_{21}>0$, there are four degenerate minima, 
\beq\label{eq: min}
(v,\theta) = \left(v_0, \frac{n\pi}{2}\right),
\eeq
and four saddle points, 
\beq\label{eq: saddle}
(v,\theta) = \left(v_1, \frac{(2n+1)\pi}{4}\right),
\eeq
where $n=0,1,2,3$, and 
\beq
v_0 = \sqrt{\frac{\mu_3^2}{\rho_1}},~\text{and}~v_1 = \sqrt{\frac{\mu_3^2}{\rho_1 + \frac{\rho_{21}}{2}}}.
\eeq
A contour plot of $V_{\chi}$ is shown in Fig.\,\ref{fig: Z4} for a fixed set of parameters. The potential has a $Z_4\simeq\mathcal{P}\times Z_2$ symmetry, with the adjacent minima connected by $\mathcal{P}$, and non-adjacent minima connected by $Z_2$. We denote the four minima $(v_R,v_L)$ as: $R+ \equiv (v_0,0), ~L+ \equiv (0,v_0), ~R- \equiv (-v_0,0), ~L- \equiv (0,-v_0)$. The desired vacuum consistent with phenomenology is the $R+$ vacuum\footnote{The $+$ or $-$ here is just a convention. We could also take $R-$ as the desired vacuum.}. After the $\mathcal{P}$-breaking PT, spatial points separated by distances larger than the correlation length $\xi$ fall into any of the four minima with equal probability. This creates a network of domain walls separating regions of distinct vacuua, each with volume $\sim \xi^3$. Past papers on DWs from LRSM have mostly discussed DWs that separate the $L$-type regions from the $R$-type regions. However, due to the $Z_4$ symmetry, two kinds of DWs are formed: 
 \begin{enumerate}
 \item $LR$ DWs, denoted by $\boxed{L\pm|R\pm}$, which separate adjacent minima, i.e., the $L\pm$ regions from the $R\pm$ regions. 
 \item $Z_2$ DWs, denoted by $\boxed{L+|L-}$ or $\boxed{R+|R-}$, separating non-adjacent minima, i.e. the $L(R)+$ regions from the $L(R)-$ regions. 
 \end{enumerate}
We will later discuss that the $Z_2$ DWs are unstable and decay into pairs of $LR$ DWs. The energy density $\mathcal{E}$ of the DW network is given by the `00' component of the energy-momentum tensor. For a single DW configuration perpendicular to the $x$-axis, separating two vacuua at $x\rightarrow \pm\infty$,
\begin{equation}
\mathcal{E} = \frac{1}{2}\left(\frac{dv_L}{dx}\right)^2 + \frac{1}{2}\left(\frac{dv_R}{dx}\right)^2 + V(v_L,v_R) + C,
\end{equation}

\noindent where $C$ is a constant chosen so that $\mathcal{E}$ vanishes at infinity. 

Due to translational symmetry, we can choose the DW profile to be centered at $x=0$. Integrating $\mathcal{E}$ along the $x$ direction yields the energy per unit area or the DW surface tension, $\sigma$,
\beq
\sigma = \int_{-\infty}^{\infty}\mathcal{E} dx.
\eeq
The kink solution interpolating between the two vacuua minimizes $\sigma$, and therefore obeys,
\beq
\frac{d}{dx}\left(\frac{\partial\mathcal{E}}{\partial (d v_i/dx)}\right) - \frac{\partial\mathcal{E}}{\partial v_i} = 0,~i\in\{L,R\}.
\eeq
We get a pair of coupled ordinary differential equations,
\bea
\frac{\partial^2 v_L}{\partial x^2} = -\mu_3^2 v_L + \rho_1 v_L^3 + \frac{\rho_2}{2} v_L v_R^2,\label{eq: kink1}\\
\frac{\partial^2 v_R}{\partial x^2} = -\mu_3^2 v_R + \rho_1 v_R^3 + \frac{\rho_2}{2} v_L^2v_R\label{eq: kink2},
\eea
which can be solved numerically using relaxation methods (see for example refs.\,\cite{Battye:2011jj,Battye:2020sxy}), with appropriate boundary conditions. To construct DW profiles, it is sufficient to consider $\boxed{L+|R+}$ and $\boxed{R-|R+}$. The boundary conditions are: \\
\noindent\textbf{Case I: $LR$ DWs} 
\beq
    \lim_{x\rightarrow -\infty} (v_R,v_L) = (0,v_0),~\lim_{x\rightarrow +\infty} (v_R,v_L) = (v_0,0). \label{eq: bc1}
\eeq

\noindent\textbf{Case II: $Z_2$ DWs} 
\beq
    \lim_{x\rightarrow -\infty} (v_R,v_L) = (-v_0,0),~ \lim_{x\rightarrow +\infty} (v_R,v_L) = (v_0,0).\label{eq: bc2}
\eeq

\begin{figure}[tbp]
    \centering
    \includegraphics[width=\textwidth]{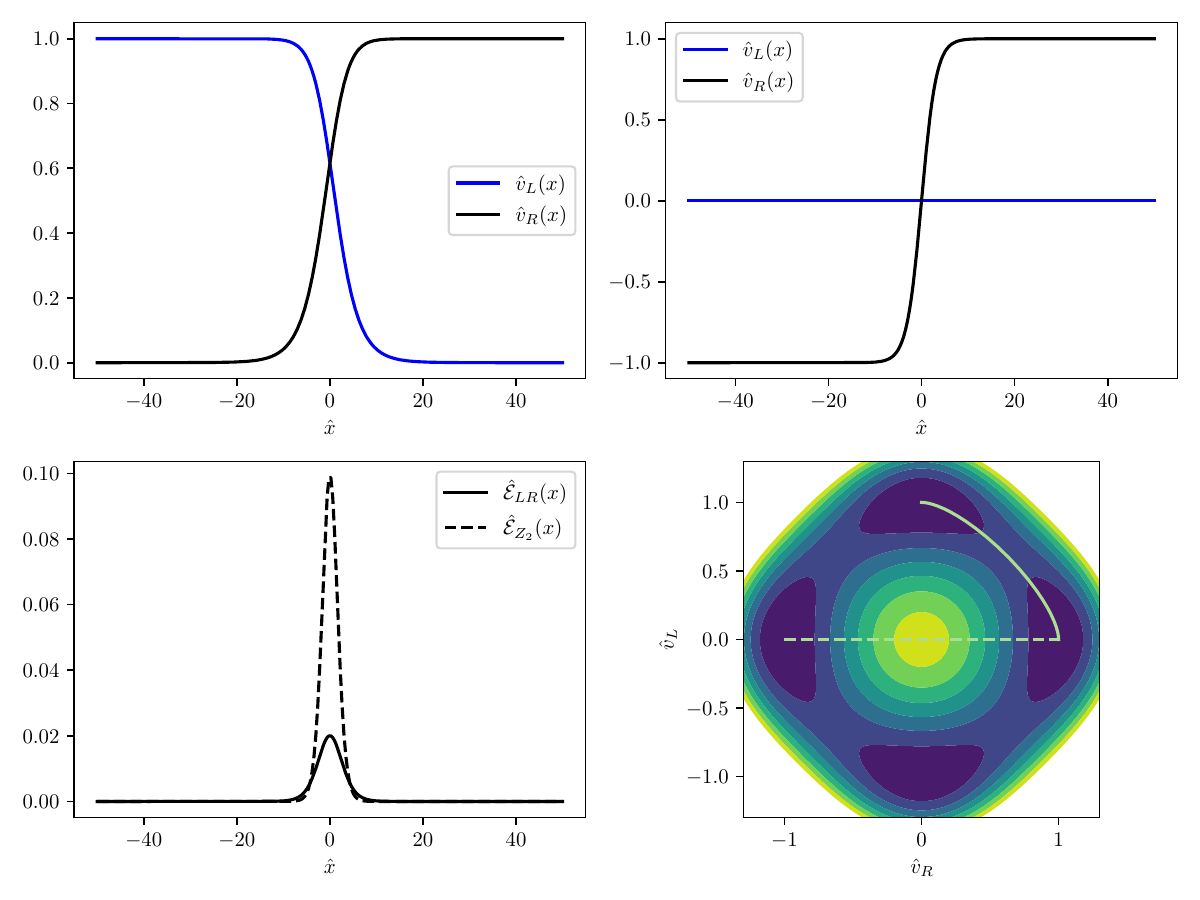}
    \caption{$LR$ (top-left panel) and $Z_2$ (top-right panel) DW profiles, for $\rho_1 = 0.2$, and $\rho_2 = 0.6$. The dimensionless energy density of the DWs is shown in the bottom-left panel. The bottom-right panel shows the $LR$ (solid line) and $Z_2$ (dashed line) DW profiles in field space.}
    \label{fig: DW_profile}
\end{figure}

\begin{figure}[tbp]
    \centering
    \includegraphics[width=\textwidth]{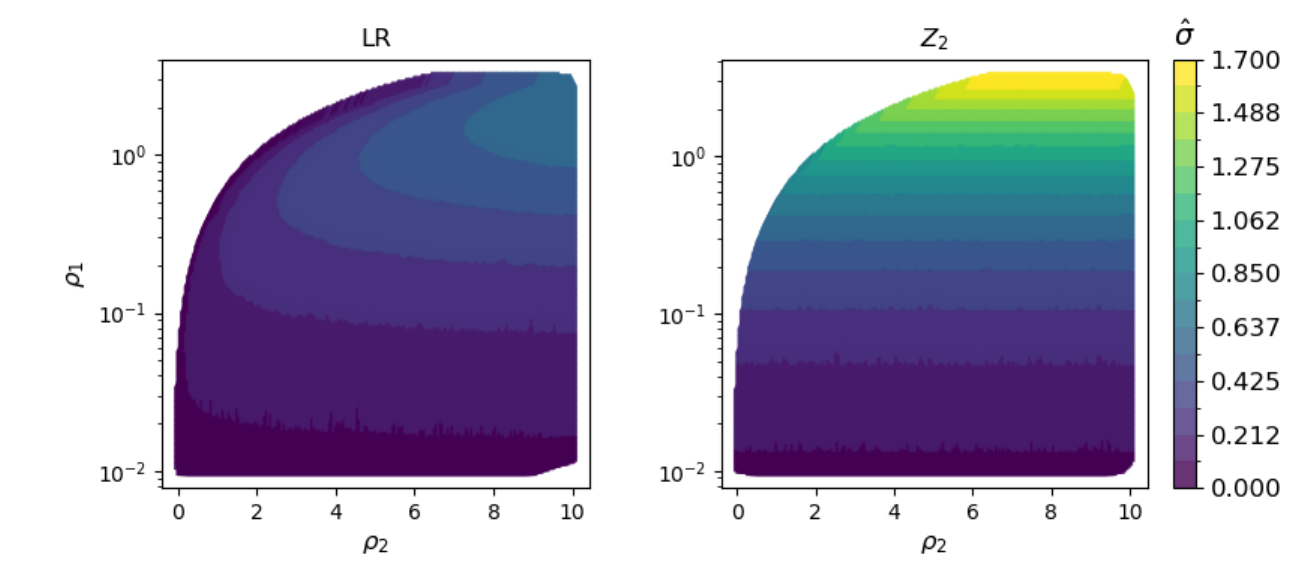}
    \caption{Parametric dependence of $\hat{\sigma}$ on $\rho_1$ and $\rho_2$ for $LR$ (left panel) and $Z_2$ (right panel) DWs. The surface tension of $Z_2$ DWs is greater than that of $LR$ DWs.}
    \label{fig: surf_ten}
\end{figure}

It is convenient to express the equations in terms of dimensionless quantities,
\beq
\hat{v}_L =  \frac{v_L}{v_0},~\hat{v}_R=  \frac{v_R}{v_0},~\hat{\mu}^2_3 =  \frac{\mu_3^2}{v_0^2} = \rho_1, ~\hat{x} = x~v_0,
\eeq
where the hatted variables are dimensionless. Similarly, we can define the dimensionless surface tension,
\beq
\hat{\s} = \frac{\s}{v_0^3} = \int_{-\infty}^{\infty} \hat{\mathcal{E}} d\hat{x},
\eeq
with $\hat{\mathcal{E}} = \frac{\mathcal{E}}{v_0^4}$ as the dimensionless energy density. We describe the procedure to obtain kink solutions in Appendix\,\ref{appendix: kink}.

In Fig.\,\ref{fig: DW_profile} we show the DW profiles for LR and $Z_2$ cases. For the chosen benchmark,
$$\hat{\s}_{LR} = 0.1363,~\hat{\s}_{Z_2} = 0.4216 \implies \frac{\hat{\s}_{Z_2}}{\hat{\s}_{LR}} \approx 3.1 \,.$$ 
The energy density of the two kinds of DWs differs significantly, as seen in the lower left panel. In field space, the $LR$ DW passes through one of the saddle points, while the $Z_2$ DW passes through the origin. Due to the greater energy density, $Z_2$ DWs are unstable and decay into $LR$ DWs, as discussed in Ref.\,\cite{Wu:2022tpe}. By adding a small shift to the initial guess of the $Z_2$ DW along the $v_L$ direction we checked that the $Z_2$ DW solution converges into a hybrid structure of two $LR$ DWs. 

\begin{figure}[tbp]
    \centering
    \includegraphics[width=0.8\textwidth]{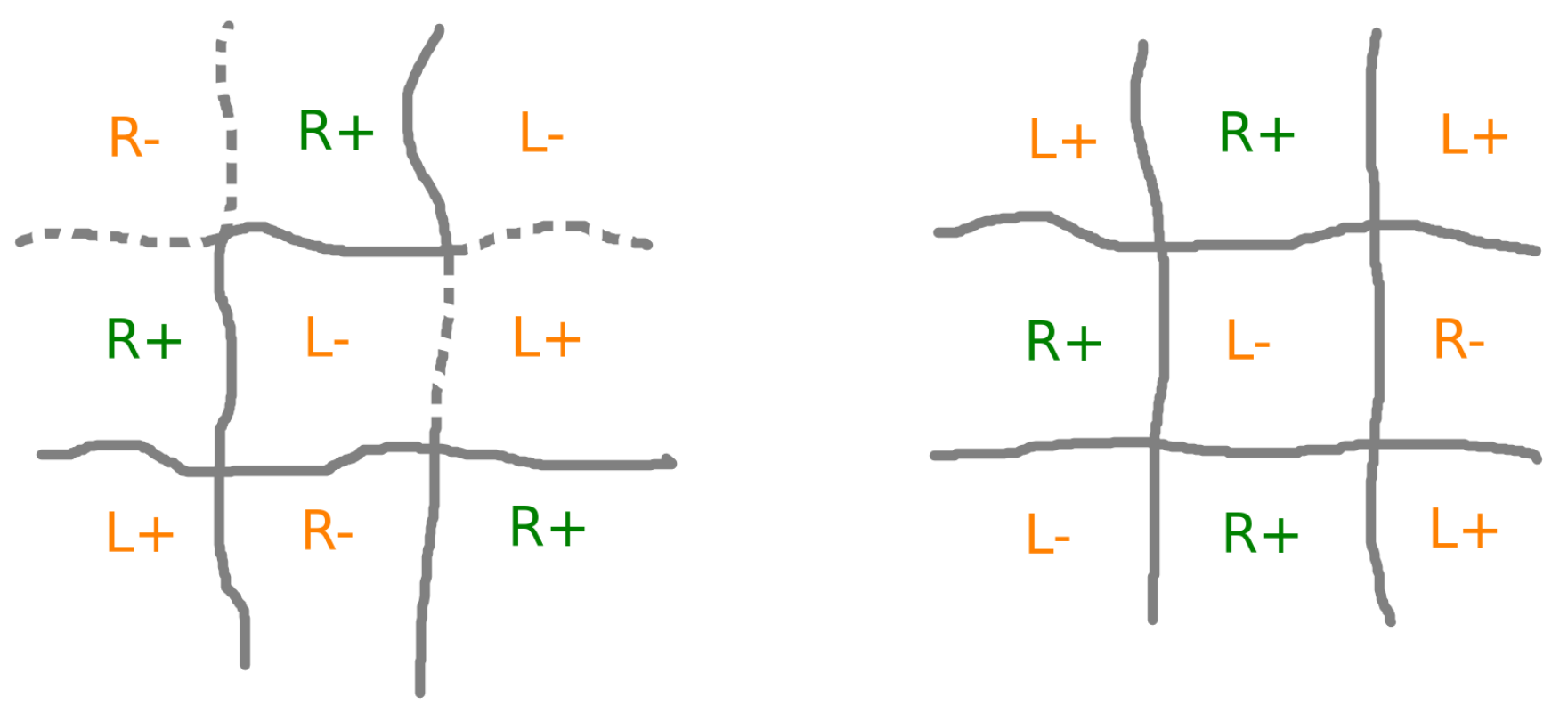}
    \caption{Left: Schematic diagram of a typical DW network showing LR (solid line) and $Z_2$ (dashed line) DWs. Right: In the stable configuration, the $Z_2$ DWs are absent and the network consists entirely of $LR$ DWs. The vacuum consistent with phenomenology, $R+$, is shown in green, while the rest are shown in orange.}
    \label{fig: DW_network}
\end{figure}

Fig.\,\ref{fig: surf_ten} shows the dependence of $\hat{\sigma}$ on the quartic couplings $\{\rho_1,\rho_2\}$, for $LR$ and $Z_2$ DWs. The surface tension of $LR$ DWs is higher for larger values of $\rho_1$ and $\rho_2$, while it is almost independent of $\rho_2$ for $Z_2$ DWs. In both cases, the overall dependence on $\rho_1$ and $\rho_2$ is weak, since the variation in $\hat{\sigma}$ in the $\rho_1-\rho_2$ plane is within an order of magnitude. $Z_2$ DWs have greater energy density than $LR$ DWs in the entire $\rho_1-\rho_2 $ plane. When $\hat{\s}_{Z_2}>2\hat{\s}_{LR}$, a $Z_2$ DW can split into two $LR$ DWs of equal area connecting adjacent minima \,\cite{Wu:2022tpe}. On the other hand if $\hat{\s}_{Z_2}<2\hat{\s}_{LR}$, then a $Z_2$ wall can split into two $LR$ DWs of smaller surface area. The most stable configuration of the DW network consists only of LR DWs after the $Z_2$ DWs have decayed away, as depicted in the right panel of Fig.\,\ref{fig: DW_network}. This explains why it is enough to focus on $LR$ DWs. This discussion applies equally to TLRSM, where the potential also obeys a $Z_4$ symmetry and has a similar vacuum structure.

\section{Gravitational waves from domain walls}\label{sec: GW}
Once formed, the DWs quickly reach a scaling regime in the absence of friction, with $\mathcal{O}(1)$ DWs per Hubble volume moving at relativistic speeds. The energy density in the scaling regime is given by\,\cite{Saikawa:2017hiv},
\beq\label{eq: rho_DW}
\rho_{\rm{DW}} = \mathcal{A} \sigma H,
\eeq
where $\mathcal{A}\sim 0.8$ is a numerical factor obtained from simulations and $H$ is the Hubble parameter. 

At time $t$, $\rho_{\rm{DW}}\propto H(t) \propto 1/t$, while the energy density of matter and radiation falls faster, implying that DWs dominate the energy density of the universe at late times. By introducing a small bias term in the potential, $V_{\rm{bias}}$, via explicit $\mathcal{P}$-breaking operators, we can lift the degeneracy of the four vacuua in such a way that the $R+$ vacuum is favored. This creates a pressure difference across the DWs, causing the domains with the preferred vacuum to grow in size. Due to the bias, the DWs eventually begin to annihilate at a temperature $T_{\rm{ann}}$, defined by the condition, $\rho_{\rm{DW}}(T_{\rm{ann}})\sim V_{\rm{bias}}$. Assuming radiation-domination\,\cite{Ferreira:2022zzo},
\beq\label{eq: Tstar}
T_{\rm{ann}} \simeq \frac{5~{\rm{MeV}}}{\sqrt{\mathcal{A}}}\left(\frac{10.75}{g_*(T_{\rm{ann}})}\right)^{\frac{1}{4}}\left(\frac{V_{\rm{bias}}^{1/4}}{10~{\rm{MeV}}}\right)^2\left(\frac{10^5~{\rm{GeV}}}{\sigma^{1/3}}\right)^{\frac{3}{2}}.
\eeq
where $g_*(T_{\rm{ann}})$ is the number of relativistic degrees of freedom at $T_{\rm{ann}}$. To accommodate cosmological constraints, DWs must annihilate before the BBN epoch, i.e. $T_{\rm{ann}}>T_{\rm{BBN}}\sim 1$\,MeV. This yields a lower bound on $V_{\rm{bias}}$,
\beq
V_{\rm{bias}} > \frac{\mathcal{A}}{25}(10~{\rm{MeV}})^4 \left(\frac{\sigma^{1/3}}{10^5~{\rm{GeV}}}\right)^3.
\eeq

Similarly, an upper bound on $V_{\rm{bias}}$ is obtained by requiring that domains of the preferred vacuum must not percolate\,\cite{Saikawa:2017hiv},
\beq\label{eq: percolation}
\frac{V_{\rm{bias}}}{V_0} > \ln\left(\frac{1-p_c}{p_c}\right),
\eeq
where $p_c=0.311$ is the critical value above which the favored vacuum percolates and $V_0$ is the height of the barrier separating the minima. For $LR$ DWs, 
\beq
V_0 = \frac{1}{4}v_0^4\left(\frac{\rho_2-2\rho_1}{\rho_2+2\rho_1}\right)\approx \frac{1}{4}v_0^4,
\eeq
where the last approximation holds when $\rho_2\gg\rho_1$. The condition of eq.\,\eqref{eq: percolation} is easily satisfied since we are interested in $V_{\rm{bias}}\ll v_0^4$. 

The bias term can be generated by introducing operators that explicitly break the $\mathcal{P}$ symmetry. Since quantum gravity effects destroy global symmetries, Planck scale-suppressed higher dimensional operators provide an elegant way to generate the bias term. Indeed, this possibility has been considered in the literature\,\cite{Lew:1993yt,Hiramatsu:2010yz,Borah:2022wdy,Barman:2023fad}. In this paper, we do not assume any particular origin of the bias term and keep $V_{\rm{bias}}$ as a free parameter.

A dimensionless parameter, $\a_*$, captures the DW energy density at annihilation, 
\bea
\a_* &=& \frac{\rho_{\rm{DW}}(T_{\rm{ann}})}{\rho_{\rm{rad}}(T_{\rm{ann}})},\nn\\
\rho_{\rm{rad}}(T) &=& \frac{\pi^2}{30}g_*(T)T^4,
\eea
Using eq.\,(\ref{eq: rho_DW}), we get\,\cite{Ferreira:2022zzo},
\beq\label{eq: alpha_star}
\alpha_* \simeq \mathcal{A}\sqrt{\frac{g_*(T_{\rm{ann}})}{10.75}}\left(\frac{\sigma^{1/3}}{10^5~{\rm{GeV}}}\right)^3\left(\frac{{10~{\rm{MeV}}}}{T_{\rm{ann}}}\right)^2 . 
\eeq
We consider the scenario where the DWs decay into standard model particles, in which case, BBN restricts $T_{\rm{ann}}\gtrsim 2.7$\,MeV \cite{Jedamzik:2006xz,PhysRevD.105.095015}. We also impose $\alpha_*<0.3$ to ensure no deviation from radiation domination\,\cite{Ferreira:2022zzo}. 
 
The relic GW spectrum is defined as, 
\beq
h^2\Omega_{\rm{GW}} = \frac{h^2}{\rho_c}\frac{\partial \rho_{\rm{GW}}}{\partial \ln f},
\eeq
where $\rho_c$ is the critical density, given by
\beq
\rho_c = \frac{3H_0^2}{8\pi G},
\eeq
and $H_0 = 100\,h~{\rm{km \,s^{-1} Mpc^{-1}}}$ is the present-day Hubble constant with $h=0.6736\pm 0.0054$\,\cite{Planck:2018vyg}, and $G$ is  Newton's gravitational constant.

GWs are produced due to DW surface oscillations\,\cite{Vilenkin:1981zs,Preskill:1991kd,PhysRevD.59.023505}, with dominant emission happening at $T=T_{\rm{ann}}$. Assuming all the GWs are produced at $T_{\rm{ann}}$, the GW spectrum is given by\,\cite{Ferreira:2022zzo,Hiramatsu:2013qaa}, 
\beq
h^2\Omega^{\rm{DW}}_{\rm{GW}}(f) \simeq 10^{-10}~ \tilde{\epsilon}_{\rm{GW}}\left(\frac{10.75}{g_*(T_{\rm{ann}})}\right)^{\frac{1}{3}}\left(\frac{\a_*}{0.01}\right)^2 S\left(\frac{f}{f_p^0}\right),
\eeq
where $\tilde{\epsilon}_{\rm{GW}} = 0.7\pm 0.4$. The peak frequency $f_p^0$ is given by,
\beq
f_p^0 \simeq 10^{-9}~{\rm{Hz}}~\left(\frac{g_*(T_{\rm{ann}})}{10.75}\right)^{\frac{1}{6}}\left(\frac{T_{\rm{ann}}}{10~{\rm{MeV}}}\right),
\eeq
and the shape function of the GW spectrum $S$ has the form,
\beq
S(x) = \frac{(\gamma+\b)^{\delta}}{\big(\b x^{-\frac{\gamma}{\delta}} + \gamma x^{\frac{\b}{\delta}}\big)^{\delta}}.
\eeq
We set $\gamma = 3$ from causality\,\cite{NANOGrav:2023hvm}, while numerical analyses determine $\delta,\beta\simeq 1$. Following\,\cite{Hiramatsu:2013qaa}, we set $\delta,\beta = 1.$ Note that, $$h^2\Omega^{\rm{DW}}_{\rm{GW}}(f_p^0)\propto \a_*^2\propto\frac{\sigma^2}{V_{\rm{bias}}}\propto \frac{v_0^6}{V_{\rm{bias}}},$$ indicating a strong dependence of the amplitude on the scale $v_0$. 

After the annihilation of DWs, the GW production stops, and $\rho_{\rm{GW}}$ redshifts like SM radiation, contributing to the number of relativistic degrees of freedom, $g_*(T)$. Around BBN temperatures, $T\simeq \mathcal{O}$(MeV), this extra contribution from GWs can be restricted by considering the limits on $\Delta N_{\rm{eff}}$ from CMB and BBN, where $\Delta N_{\rm{eff}} = N_{\nu}-3$, and $N_{\nu}$ is the effective number of light neutrino species at BBN. The upper bound on the GW amplitude is\,\cite{Maggiore:1999vm,Caprini:2018mtu}, 
\beq
h^2\Omega_{\rm{GW}}\lesssim 5.6\times 10^{-6}\Delta N_{\rm{eff}}.
\eeq
The existing limit on $\Delta N_{\rm{eff}}$ from Planck is, $\Delta N_{\rm{eff}}\lesssim 0.28$ at $95\%$ confidence level. Upcoming CMB experiments will be able to probe smaller values of $\Delta N_{\rm{eff}}$\,\cite{Planck:2018vyg}.

\section{Results}\label{sec: results}
\subsection{MCMC analysis}
We use the NG15 data to carry out the Bayesian analysis. Given PTA data $\mathcal{D}$, a hypothesis $\mathcal{H}$, and parameters $\Theta$, we use the posterior distribution $P (\Theta|\mathcal{D},\mathcal{H})$, reconstructed from MCMC analysis, to identify best-fit parameter ranges and set upper limits on them. We consider two hypotheses: (i) $\mathcal{H}_1$: the DLRSM DW model assuming the SMBH background is negligible, and (ii) $\mathcal{H}_2$: the DLRSM DW$+$SMBHB model, where the GW contribution from DLRSM DWs is combined with the contribution from SMBHBs. The DLRSM parameters of interest are,
$$\Theta_1\equiv\{v_0,\rho_1,\rho_2,V_{\rm{bias}}\}.$$

For $\mathcal{H}_1$, we first obtain the functional dependence of $\hat{\sigma}(\rho_1,\rho_2)$ by interpolating the numerical values shown in Fig.\,\ref{fig: surf_ten}, so that the surface tension is obtained as, $\sigma(v_0,\rho_1,\rho_2) = \hat{\sigma}(\rho_1,\rho_2) v_0^3$. Next, we calculate $T_{\rm{ann}}$  according to eq.\,\eqref{eq: Tstar}, while $\a_*$ is calculated using eq.\,\eqref{eq: alpha_star}. The constraints $T_{\rm{ann}}\gtrsim 2.7$\,MeV and $\a_*<0.3$,  are imposed as discussed in the previous section. Thus we obtain the GW spectrum in terms of DLRSM parameters
$$\mathcal{H}_1:~~~ h^2\Omega^{\rm{DW}}_{\rm{GW}}(f;v_0,\rho_1,\rho_2,V_{\rm{bias}}).$$

For $\mathcal{H}_2$, we superimpose the contribution of SMBHBs with the DLRSM DW contribution. For low frequencies, $f\ll 1\,\rm{year}^{-1}$, the SMBHB spectrum is given by a simple power law\,\cite{1995ApJ...446..543R,Jaffe_2003,Wyithe_2003,Sesana_2004,Burke-Spolaor2019}, 
\beq\label{eq: bhb_gw}
h^2\Omega^{\rm{BHB}}_{\rm{GW}}(f) = \frac{2\pi^2h^2 A^2_{\rm{BHB}}}{3H_0^2} \left(\frac{f}{f_{\rm{yr}}}\right)^{5-\gamma_{\rm{BHB}}} f_{\rm{yr}}^2,
\eeq
where $f_{\rm{yr}} = 1\, \rm{year}^{-1} = 3.17\times 10^{-8}$\,Hz. The spectrum falls off rapidly at larger frequencies, $f\gg 1\,\rm{year}^{-1}$. If the orbital evolution of the binaries is purely driven by GW emission, the parameter $\gamma_{\rm{BHB}}$ takes the value, $ \gamma_{\rm{BHB}} = 13/3$. To account for environmental effects $\gamma_{\rm{BHB}}$ is taken as a free parameter, along with $A_{\rm{BHB}}$, so that the parameter set is,
$$\Theta_2\equiv\{v_0,\rho_1,\rho_2,V_{\rm{bias}},A_{\rm{BHB}},\gamma_{\rm{BHB}}\}.$$
The GW signal is hypothesised as,
$$\mathcal{H}_2:~~~ h^2\Omega^{\rm{DW}}_{\rm{GW}}(f;v_0,\rho_1,\rho_2,V_{\rm{bias}}) + h^2\Omega^{\rm{BHB}}_{\rm{GW}}(f;A_{\rm{BHB}},\gamma_{\rm{bhb}}).$$

\begin{figure}[tbp]
    \centering
    \includegraphics[width=.8\textwidth]{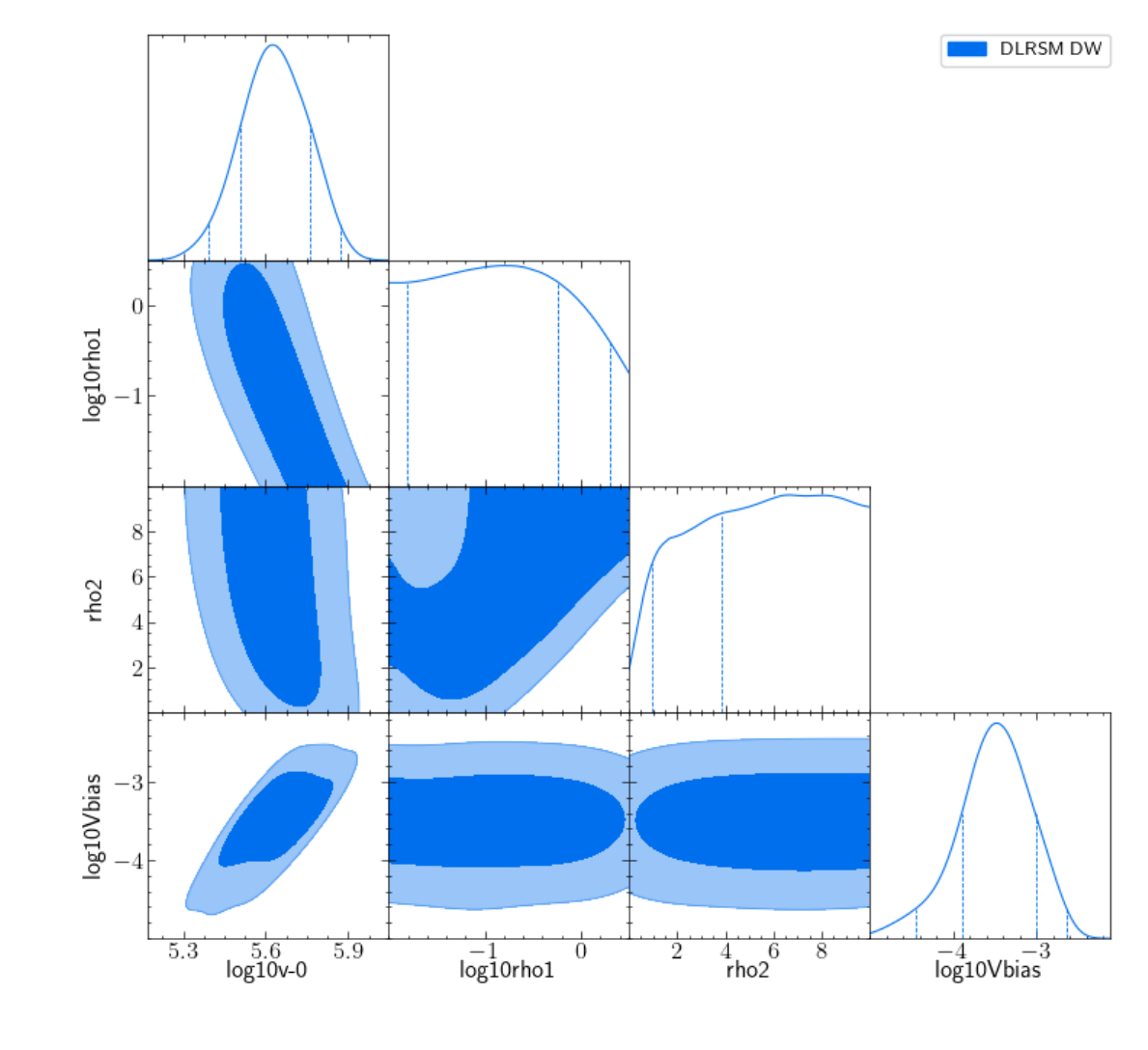}
    \caption{The posterior probability distribution of the DLRSM DW fit to the NG15 data.\label{fig: MCMC}}
\end{figure}

\begin{figure}[tbp]
    \centering
    \includegraphics[width=\textwidth]{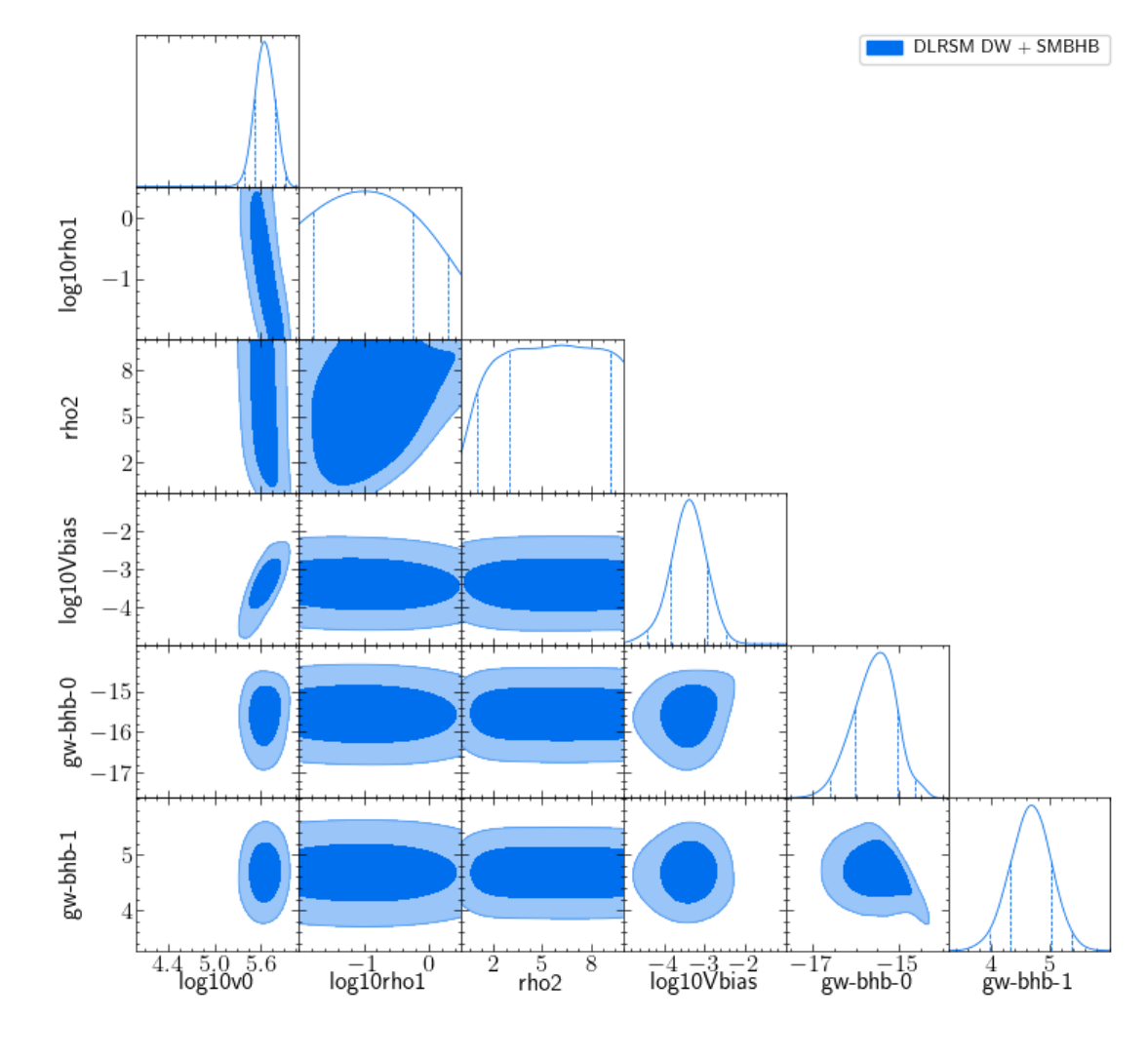}
    \caption{Same as fig.\,\ref{fig: MCMC}, but also including the contribution from SMBHB.}
    \label{fig: MCMC_SMBHB}
\end{figure}

The Bayesian analysis is implemented using the \texttt{PTArcade} package\,\cite{andrea:mitridate_2023,Mitridate:2023oar}, which is a wrapper for the \texttt{ENTERPRISE} code\,\cite{ellis_2020_4059815}, and incorporates PTA data. For the DLRSM DW model, we sample the parameters $v_0,~V_{\rm{bias}}$ and $\rho_1$ from a log-uniform distribution, and $\rho_2$ from a uniform distribution. The parameter ranges for the priors are given in the second column of Table\,\ref{table: priors}. The SMBHB parameters ($\log_{10}A_{\rm{BHB}},\gamma_{\rm{BHB}}$) are sampled from a normal bivariate distribution taken from Ref.\,\cite{NANOGrav:2023hvm}.

The posteriors for DLRSM DWs are shown in Fig.\,\ref{fig: MCMC}. The $\mathcal{P}$-breaking scale $v_0$ and the bias term $V_{\rm{bias}}$ are constrained in a narrow range, as indicated by the closed contours for the $68\%$ and $95\%$ credible intervals. On the other hand, there is a weak dependence on the quartic couplings $\rho_1$ and $\rho_2$. Note also that $\rho_1\ll\rho_2$ in the $68\%$ and $95\%$ credible regions. A positive correlation is observed between $v_0$ and $V_{\rm{bias}}$, which is because, for a given scale $v_0$, we can always find an appropriate value of $V_{\rm{bias}}$, which best explains the data. In Fig.\,\ref{fig: MCMC_SMBHB}, we show the posterior distribution for the DLRSM DW$+$SMBHB model, with the labels gw-bhb-0 and gw-bhb-1 denoting the SMBHB parameters $\log_{10}A_{\rm{BHB}}$ and $\gamma_{\rm{BHB}}$ respectively. The DLRSM parameters follow a similar distribution as in the previous case, and the other two parameters take the maximum posterior values $(\log_{10}A_{\rm{BHB}}, \gamma_{\rm{BHB}}) = (-15.44,4.69)$. The predicted spectral index, $\gamma_{\rm{BHB}} = 13/3$, lies just outside the $68\%$ credible interval. The likelihood ratio of the DLRSM DW relative to the the DLRSM DW$+$SMBHB model corresponds to, $-2 \Delta\ln l_{\rm{max}} = -1.2$, indicating that the pure DLRSM DW model is slightly favored.

Table\,\ref{table: priors} summarizes the prior ranges, maximum posteriors, and $68\%$ credible intervals for all the parameters in the two scenarios. In particular, the maximum posterior values of $v_0$ and $V_{\rm{bias}}$ for DLRSM DW model are,
\beq
v_0 = 4.36\times 10^5\,\rm{GeV}, ~ V_{\rm{bias}} = 3.31\times 10^{-4}\,\rm{GeV}^4.
\eeq 

\begin{figure}[tbp]
    \centering
    \includegraphics[width=\textwidth]{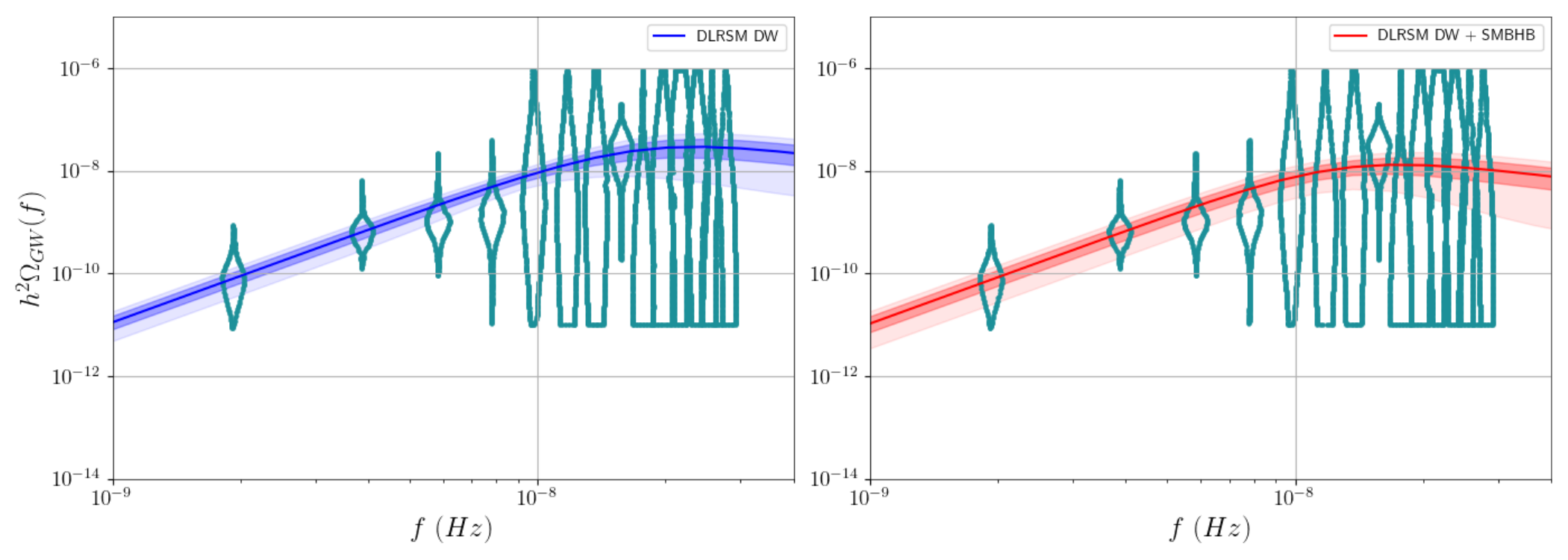}
    \caption{Median GW spectra for DLRSM DWs, along with their $68\%$ and $95\%$ posterior envelopes.}
    \label{fig: GW spectrum}
\end{figure}

\begin{table}[tbp]
\begin{center}
\def\arraystretch{1.5}
\setlength{\tabcolsep}{8pt}
\resizebox{\textwidth}{!}{
\begin{tabular}{c c c c c c}
\hline
\multicolumn{1}{c}{\multirow{2}{*}{\textbf{Parameter}}} & \multicolumn{1}{c}{\multirow{2}{*}{\textbf{Uniform prior range}}} & \multicolumn{2}{c}{\multirow{2}{*}{\textbf{Maximum Posterior}}} & \multicolumn{2}{c}{\multirow{2}{*}{\textbf{$68\%$ credible interval}}}  \\ 
\multicolumn{1}{c}{}  & & $\mathcal{H}_1$ & $\mathcal{H}_2$  & $\mathcal{H}_1$ & $\mathcal{H}_2$    \\ \hline
$\log_{10}~v_0/\rm{GeV}$ &  $[4,8]$ & $5.63$ & $5.66$&  $[5.52,5.77]$ & $[5.52,5.79]$ \\
$\log_{10}~V_{\rm{bias}}/\rm{GeV}^4$ & $[-5,-1]$ & $-3.47$ & $-3.37$ & $[-3.90,-3.05]$ & $[-3.82,-2.91]$\\  
$\log_{10}\rho_1 $    & $[-2.5,0.5]$ & $-0.77$ & $-1.00$ & $[-1.90,-0.35]$ & $[-1.76,-0.23]$\\  
$\rho_2 $    & $[0,10]$ & $6.55$ & $6.50$ & $[2.88,9.16]$ & $[3.03,9.23]$\\   
$\log_{10}A_{\rm{BHB}}$ & - & - & $-15.44$ & - & $[-15.97,-15.02]$\\
$\gamma_{\rm{BHB}}$     & - & - & $4.69$  & - & $[4.35,5.04]$\\
\hline
\end{tabular}}
\end{center}
\caption{\label{table: priors} Priors, along with the maximum posterior values and $68\%$ credible intervals for the parameters, for the two hypotheses considered.
}
\end{table}

\subsection{Detection prospects}
We now discuss the prospects of detecting the GW signal using the best-fit parameter values at upcoming GW observatories. In Fig.\,\ref{fig: GW spectrum} we show the median GW spectra fitted to the NG15 data along with $1\sigma$ and $2\sigma$ confidence intervals, for the DLRSM DW model (left panel), and DLRSM DW combined with SMBHB (right panel). The green violins correspond to the posterior distribution of $\Omega_{\rm{GW}}$ in 14 frequency bins, predicted by the NG15 data\,\cite{NANOGrav:2023gor}. With more data, the posteriors would narrow down, enabling a more precise parameter estimation in the future. 

Fig.\,\ref{fig: spectra}  shows the median GW spectra for the two models for a wider range of frequencies, with the power-law integrated sensitivity curves (PLISCs)\,\cite{Thrane:2013oya} of upcoming GW detectors, SKA\,\cite{Weltman:2018zrl}, $\mu$Ares\,\cite{Sesana:2019vho}, LISA\,\cite{LISA:2017pwj}, BBO\,\cite{Corbin:2005ny}, FP-DECIGO\,\cite{Seto:2001qf}, CE\,\cite{LIGOScientific:2016wof}, and ET\,\cite{Punturo:2010zz}. The shaded grey region shows the region excluded by the $\Delta N_{\rm{eff}}$ bound coming from Planck data. Except $\mu$Ares, the PLISCs of all upcoming detectors are taken from Ref.\,\cite{schmitz_2020_3689582}, using the threshold signal-to-noise ratio (SNR) of 1, and time of observation, $\tau=1$\,year, while $\tau=20$\, years for SKA. The PLISC for $\mu$Ares is calculated for threshold SNR of 10, and $\tau=7$\,years. If the GW spectrum results from the DLRSM DW model, considered with or without SMBHBs, it would be observed by future observatories, particularly $\mu$Ares, LISA, FP-DECIGO, and BBO, with a high SNR. The signal will also be confirmed by upcoming PTAs such as SKA. If the $\mathcal{P}$-breaking PT is first-order, one would observe a double-peaked GW spectrum, with the additional higher frequency peak coming from FOPT, as discussed in Ref.\,\cite{Karmakar:2023ixo}. For $v_0\gtrsim 10^5$\,GeV, the peak from FOPT would be detected by CE an ET. 

In DLRSM, the extra heavy degrees of freedom including neutral $CP$-even scalars $H_{1,2,3}$, $CP$-odd scalars $A_{1,2}$, charged scalars $H_{1,2}^{\pm}$ and gauge bosons $Z_2, W_2^{\pm}$, all have $\mathcal{O}(v_0)$ masses. Due to the high parity-breaking scale favored by the NG15, $v_0\gtrsim 10^5$\,GeV, the prospects of detecting these heavy particles at upcoming collider experiments are weak.

\section{Discussion}\label{sec: discussion}
After the 15-year NANOGrav dataset analysis reported convincing evidence of a low-frequency GW background, several works have compared the possible models to explain it. In addition to the standard interpretation of GWs produced by SMBHBs, many cosmological models have been proposed. One such model is the DW model, in which the GW background is due to a network of DWs in the early universe.  The formation of DWs requires the existence of a discrete symmetry which is spontaneously broken in a phase transition. In LRSMs, the discrete parity symmetry can be elegantly incorporated to explain the parity violation observed in SM via spontaneous symmetry breaking. The parity-breaking PT in LRSM can give rise to DWs. Since LRSMs typically require the scale of parity-breaking, $v_0$, to be large compared to the EW scale, the DW surface tension, $\sigma\propto v_0^3$, can be made large enough to explain the NG15 result. In this work, we considered the DWs of the parity symmetric DLRSM as the source of the NG15 signal.

Earlier discussions on DWs in LRSM mostly focused on TLRSM and reported the scale of $\mathcal{P}$-breaking required to explain the PTA signal using benchmarks obtained from order of magnitude estimates. In this paper, for DLRSM, we found the explicit parameter dependence of the surface tension and carried out a Bayesian analysis to constrain model parameters. Due to the $Z_4$ symmetry of the DLRSM potential, two kinds of DWs, i.e. $LR$ and $Z_2$ DWs are formed. The surface tension of the $Z_2$ walls is higher than that of $LR$ DWs. Earlier works on DWs in LRSM mainly focused on $LR$ DWs and did not discuss the fate of $Z_2$ DWs in detail. We argued that the $Z_2$ DWs are unstable and decay into $LR$ DWs, thus providing a rationale for considering only $LR$ DWs. The arguments presented here can also be applied to TLRSM. The DW surface tension depends weakly on the quartic couplings $\rho_1,\rho_2$, and the primary dependence is on $v_0$. The DW network must annihilate before the epoch of BBN to respect cosmological constraints, which can be achieved via explicit parity-breaking terms in the potential, resulting in a bias, $V_{\rm{bias}}$. While such operators can be motivated by quantum gravity and grand unified theories, we considered $V_{\rm{bias}}$ as a free parameter. 

\begin{figure}[tbp]
    \centering
    \includegraphics[width=0.8\textwidth]{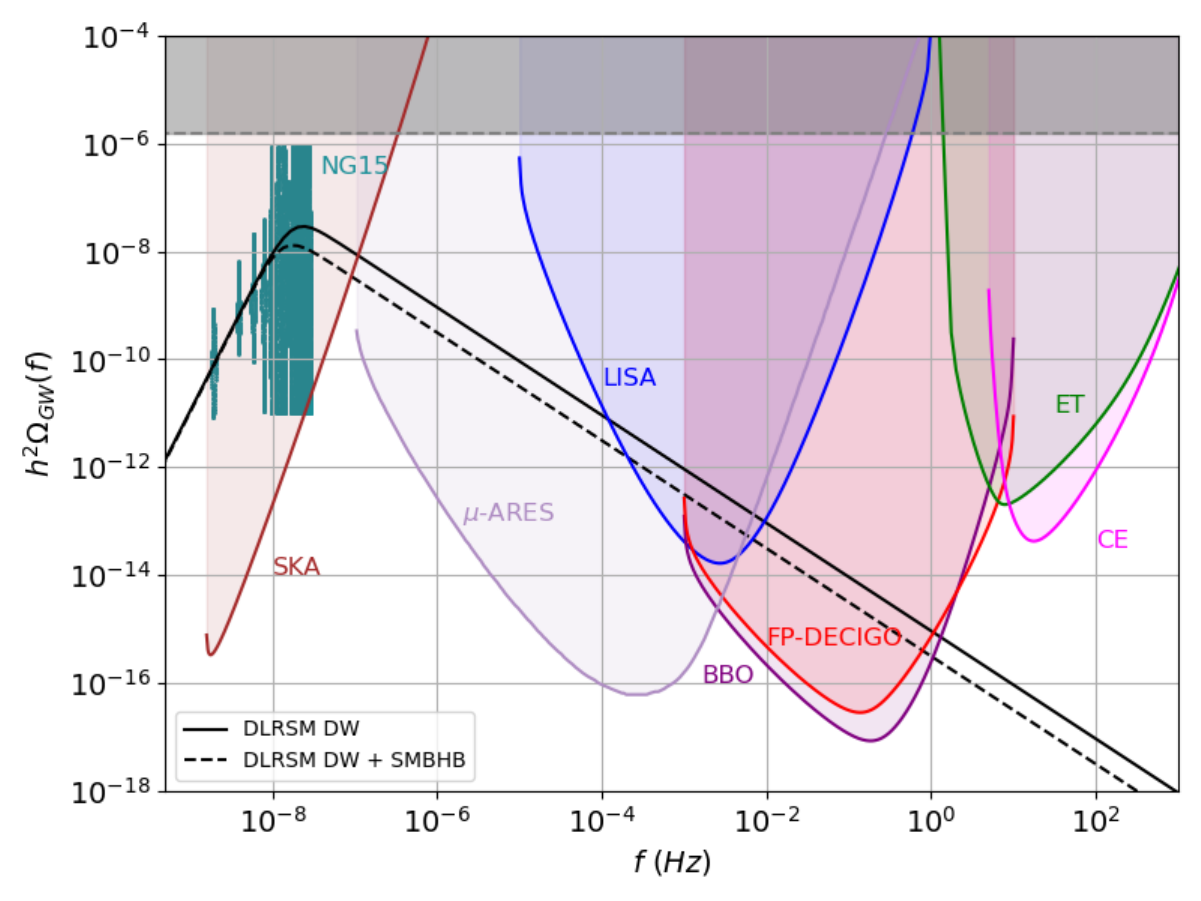}
    \caption{Median GW spectrum for DLRSM DW (solid black) and DLRSM DWs$+$SMBHB model (dashed black) along with sensitivity curves for upcoming GW detectors. The shaded grey region is excluded by the Planck $\Delta N_{\rm{eff}}$ bound.}
    \label{fig: spectra}
\end{figure}

For the Bayesian analysis, we considered the case where the NG15 signal is entirely explained by DLRSM DWs and the case where the SMBHB contribution is also included. The maximum posterior values of the parameters and the $68\%$ credible intervals are summarized in Table\,\ref{table: priors}. The strong dependence of the peak amplitude of the GW spectrum on $\sigma$, and $V_{\rm{bias}}$, i.e. $\Omega^{DW}_{\rm{GW}}(f_{\rm{peak}})\propto\sigma^2/V_{\rm{bias}}$, results in a tight constraint on $v_0$ and $V_{\rm{bias}}$, as reflected by the respective $68\%$ credible intervals. On the other hand, the quartic couplings $\rho_1$ and $\rho_2$ are less constrained. For the DLRSM DW$+$DW case, the spectral index $\gamma_{\rm{BHB}}$ lies outside the $68\%$ credible interval, indicating a tension with the NG15 results. 

The median GW spectra are presented in Fig.\,\ref{fig: GW spectrum} and Fig.\,\ref{fig: spectra}, which show a good agreement with the violins of NG15 data. If the signal is due to DWs, future GW observatories such as $\mu$Ares, LISA, BBO, and FP-DECIGO would also observe a GW background at higher frequencies. Moreover, if the parity-breaking PT is first-order, it would give rise to a double-peaked spectrum, with the higher frequency peak observable by ET and CE. 

In this paper, we have neglected the effect of friction on the DWs from the thermal plasma, which could dampen the GW signal. Since the couplings of the fields constituting the wall with the SM fields are small, the friction is expected to be negligible. In addition, the GW spectrum from SMBHBs is modeled assuming the binaries lose their energy entirely from GW production. Since current observations numerical simulations have a large uncertainty in the spectral shape, the power-law given in eq.\,\eqref{eq: bhb_gw} serves as a reasonable approximation\,\cite{NANOGrav:2023hvm}.

\section{Acknowledgements}
 DR thanks Subhendu Rakshit, Siddhartha Karmakar, and Suman Majumdar for helpful discussions. This work is supported by DST, via SERB Grants no.\,MTR/2019/000997 and\\ no.\,CRG/2019/002354. 

\appendix
\section{Kink solutions}\label{appendix: kink}
The dimensionless energy density is given by,
\bea
\hat{\mathcal{E}} = \frac{\mathcal{E}}{v_0^4} = \frac{1}{2}\left(\frac{d\hat{v}_L}{dx}\right)^2 + \frac{1}{2}\left(\frac{d\hat{v}_R}{dx}\right)^2 + \hat{V}(\hat{v}_L,\hat{v}_R) + \hat{C}.
\eea
We can rescale eq.\,(\ref{eq: kink1}) and eq.\,(\ref{eq: kink2}) as,
\bea
\frac{\partial^2 \hat{v}_L}{\partial \hat{x}^2} &=&  - \hat{\mu}_3^2 \hat{v}_L +\rho_1 \hat{v}_L^3 + \frac{\rho_2}{2} \hat{v}_L\hat{v}_R^2\label{eq: kink1_app}\\
\frac{\partial^2 \hat{v}_R}{\partial \hat{x}^2} &=&  - \hat{\mu}_3^2 \hat{v}_R +\rho_1 \hat{v}_R^3 + \frac{\rho_2}{2} \hat{v}_L^2\hat{v}_R.\label{eq: kink2_app}
\eea
In relaxation methods, a fictitious `time' variable, $\hat{t}$ is introduced, and the above equations are written as, 
\bea
\frac{\partial \hat{v}_L}{\partial \hat{t}} &=& \frac{\partial^2 \hat{v}_L}{\partial \hat{x}^2} + \hat{\mu}_3^2 \hat{v}_L -\rho_1 \hat{v}_L^3 - \frac{\rho_2}{2} \hat{v}_L\hat{v}_R^2\\
\frac{\partial \hat{v}_R}{\partial \hat{t}} &=& \frac{\partial^2 \hat{v}_R}{\partial \hat{x}^2} + \hat{\mu}_3^2 \hat{v}_R -\rho_1 \hat{v}_R^3 - \frac{\rho_2}{2} \hat{v}_L^2\hat{v}_R.
\eea
Eq.\,(\ref{eq: kink1_app}) and  eq.\,(\ref{eq: kink2_app}) are recovered in the limit $\frac{\partial }{\partial \hat{t}}\hat{v}_{L,R}\rightarrow 0$. 
We discretize the spatial and temporal coordinates with step size $\Delta \hat{x}$ and $\Delta \hat{t}$ respectively and express the derivatives in terms of second-order finite differences. The solution converges if $\Delta \hat{t} \leq \Delta\hat{x}^2/2$.  

For numerical purposes, we approximate spatial infinity by a value $R=50$ and linearly interpolate the boundary conditions eq.\,(\ref{eq: bc1}) and eq.\,(\ref{eq: bc2}), for the initial guess of $LR$ and $Z_2$ solutions. 

\bibliographystyle{unsrt}
\bibliography{citation.bib}
\end{document}